\documentclass [a4paper, 12pt] {article}
\usepackage[cp866]{inputenc}
\usepackage{amsfonts,amsmath,amssymb,amscd,euscript,epsfig}
\usepackage[english,russian]{babel}
\normalbaselines
\catcode `\@=11
\@addtoreset{equation}{section}

\def\section{\@startsection {section}{1}{\z@}{-3.5ex plus -1ex minus
     -.2ex}{2.3ex plus .2ex}{\normalsize\bf}}
\def\subsection{\@startsection{subsection}{2}{\z@}{-3.25ex plus -1ex
minus
 -.2ex}{1.5ex plus .2ex}{\normalsize\bf}}

\def\thebibliography#1{\section*{References}
\list
  {[\arabic{enumi}]}{\settowidth\labelwidth{[#1]}\leftmargin\labelwidth
  \advance\leftmargin\labelsep
  \usecounter{enumi}}
  \def\newblock{\hskip .11em plus .33em minus -.07em}
  \sloppy
  \sfcode`\.=1000\relax}
 
\catcode `\@=12
\newcommand{\be}{\begin{equation}\label}
\newcommand{\ee}{\end{equation}}

\newcommand{\p}{\prime}
\newcommand{\bib}{\bibitem}

\newcommand{\mb}{\mathbb}      
\newcommand{\wh}{\widehat}      

\begin{document}

\begin{center}

{\bf Dimerous Electron and Quantum Interference \\
beyond the Probability Amplitude Paradigm}

\vskip2mm

{\bf Vladimir V. Kassandrov}

\vskip2mm

{\it Institute of Gravitation and Cosmology, \\
Peoples' Friendship University of Russia, Moscow, Russia}

\end{center}

\vskip2cm
\noindent
{\bf Abstract.}  {\footnotesize We generalize the formerly proposed 
relationship between a special complex geometry (originating from the 
structure of biquaternion algebra) and induced real geometry 
of (extended) space-time. The primordial dynamics in complex 
space allows for a new realization of the ``one electron Universe'' 
of Wheeler-Feynman (the so called ``ensemble of duplicons'') and 
leads to a radical concept of ``dimerous'' (consisting of two 
identical matter pre-elements, ``duplicons'') electron. Using this 
concept, together with an additional phase-like invariant (arising from the 
complex pre-geometry), we manage to give a visual classical 
explanation for quantum interference phenomena and, in 
particular, for the canonical two-slit experiment. Fundamental 
relativistic condition of quantum interference generalizing the 
de Broglie relationship is obtained, and an experimentally verifiable 
distinction in predictions of quantum theory and presented  
algebrodynamical scheme is established.}

\section{Introduction. Algebrodynamics in the primordial complex space}

In our recent papers~\cite{Dupl,Mink,Yad} we have been elaborating 
the concept that primordial physical dynamics takes, in fact, place in 
the {\it complexified space-time} ${\mb C}^3$, an invariant subspace 
of the vector space of {\it biquaternion algebra} $\mb B$. It was  
assumed that structure of the latter entirely encodes both 
the geometry of physical 
space-time and the dynamics of physical fields and particles. 
Corresponding approach originating from the monograph~\cite{AD}~\footnote 
{References to the older works on algebrodynamics can be found therein, to 
the later ones -- in the review~\cite{Qanalys}}, as well as from 
the ideas of W. Hamilton, C. Lanczos, D. Hestenes et al., has been 
called the {\it algebrodynamics}. 

As to the {\it real} physical geometry, it is determined by the {\it modulus 
part} $s^2$ of the principal complex invariant $\sigma$ of the 
{\it automorphism group} $SO(3,{\mb C})$ of $\mb B$ algebra
\be{sigma}
\sigma:= {\bf z} \cdot {\bf z}, ~~~{\bf z} \in {\mb C}^3  
\ee
which, though non-negatively definite, can be equivalently represented in a 
remarkable Minkowski-like form~\cite{Mink,Feudorov}:
\be{Minkinv}
s^2: = \sigma \sigma^* = ({\bf z})^2 ({\bf z}^*)^2  \equiv 
({\bf z} \cdot {\bf z}^*)^2 -\vert \imath {\bf z} \times {\bf z}^* \vert^2
=t^2 - {\bf r}^2 \ge 0, 
\ee
with 
\be{effect}
t:= {\bf z} \cdot {\bf z}^*,~~~{\bf r}:= \imath {\bf z} \times {\bf z}^*
\ee
being effective time-like and space-like coordinates of the induced 
real geometry, respectively. Under the $SO(3,{\mb C})$-automorphisms quantities $t$ and ${\bf r}$ 
transform in a Lorentz-like way (see the details in~\cite{Mink}). Thus, the   
(macro)geometry invariantly induced by the structure of biquaternions  
actually corresponds to the {\it causal part of the Minkowski space} $\bf M$, 
with invariant $s^2$ in the role of a (necessarily non-negative) 
Minkowski interval. 

Moreover, the compact {\it phase part} of complex invariant $\sigma$ 
gives rise to an {\it $SO(3,{\mb C})$-invariant geometrical phase} $\alpha$
``attached'' to any point of the effective space $\bf M$. This phase 
turns out to be responsible for geometric explanation of wave properties of 
matter~\cite{Yad}, see also below. 

As to physical fields, these originate as the analogue of 
complex analytical functions generalized to the case of $\mb B$ algebra. 
Because of non-commutativity of $\mb B$, resultant generalization of 
the {\it Cauchy-Riemann analyticity conditions} turns out to be 
{\it nonlinear} and represents itself the equations of unique 
fundamental field, the {\it biquaternionic field}, which is thus 
{\it self-interacting} and possesses, moreover, a natural {\it twistor 
(2-spinor) structure}. 
Gauge (complex Maxwell and $SL(2,{\mb C})$ Yang-Mills) fields also find their 
place in the scheme. For exposition of non-commutative analysis (over 
quaternion-like algebras) and associated physical fields we refer the reader 
to the  review~\cite{Qanalys}. 

Finally, in the framework of algebrodynamics, particles can be naturally 
identified with various types of singularities of corresponding 
``$\mb B$-meromorphic'' functions-fields. Due to the presence of twistor 
structure, such a function gives rise to a light-like geometrical structure, 
namely, to a {\it shear-free congruence of rectilinear null rays}, both in 
the primordial complex and in the induced real space-time~\cite{Newman}. 
Within such a picture, particles (extended or point-like) correspond to 
{\it caustics,  cusps} or {\it focal lines} of the above congruence~\cite{Dupl}. 
Condition for caustic locus etc. plays the role 
of equation of particles' motion and, at the same time, determines an  
instantaneous distribution of particle-like formations in space. 

In this way, 
general physical picture arising in the framework of algebrodynamics seems 
to be self-consistent and closed. It follows only from the internal properties 
of biquaternions and $\mb B$ analytical functions so that none additional  
canonical structures (e.g. Lagrangian, `` external'' symmetry group, 
quantization rules etc.) are introduced ``by hands'' in the scheme. 

The goal of the below presented paper is to elaborate further the principal  
features of algebraic kinematics (dynamics) of particles-singularities in 
the primordial complex $\mb B$ space and its ``image'' as it looks like in  
the associated real space-time $\bf M$. In particular, in Sec.2 we specify 
(generalize) the above described relationship between the primordial complex 
$\mb B$ geometry and the invariantly associated Minkowski-like real space-time.  
In Sec.3, we briefly review the most interesting and intriguing features of  
algebrodynamics in the primordial $\mb B$ space, in particular, the concepts 
of {\it duplicons} and of {\it dimerous ``electrons''} formerly introduced 
in~\cite{Dupl} and~\cite{Yad}, respectively. In Sec.4, we make use 
of these concepts for alternative, purely classical explanation of the 
{\it quantum interference phenomenon} preliminarily presented in~\cite{Yad}. 
Sec.5 contains some final remarks on perspectives and status of 
the algebrodynamical theory.

\section{Symmetries of biquaternion algebra and the induced real space-time}   

Apart from invariant (\ref{sigma}), there is also the zeroth component $z^0$ of 
a biquaternion $Z$ which is also invariant under the $SO(3,{\mb C})$ 
automorphisms of $\mb B$ and should, generically, contribute to the effective 
real geometry. As to physical motivations and consequences of the subsequent 
generalization of the induced geometry, they will become clear afterwards.

Specifically, in the canonical matrix representation of an element $Z\in \mb B$ of 
the biquaternion algebra
\be{matrep}
Z = \begin{pmatrix}
u & w \\
p & v 
\end{pmatrix}
= \begin{pmatrix}
z^0 +z^3 & z^1 - \imath z^2 \\
z^1+\imath z^2 & z^0 - z^3 
\end{pmatrix}
\ee
where $\{u,w,p,v\}\in {\mb C},~~\{z^0,z^a\}\in {\mb C}, ~a=1,2,3$, 
principal complex invariant $\Sigma\in {\mb C}$ corresponds to the {\it determinant} 
\be{det}
\Sigma:= \det Z = (z^0)^2 - {\bf z}^2
\ee
whose modulus part $S^2:=\Sigma \Sigma^*$ is responsible for the real 
``macrogeometry'' related to the full structure of vector space of $\mb B$. 
Making use of the evident identity (generalizing (\ref{Minkinv})): 
\be{idnt}
S^2= (\det Z)(\det Z)^* = \det ZZ^+ \equiv \det X, 
\ee
one arrives again at the Minkowski-like geometry with effective space-time 
coodinates $T,{\bf R}$ forming, as usual, the structure of a 
{\it Hermitean}  matrix $X:=ZZ^+$:
\be{Hermit}
X \equiv X^+ = ZZ^+ = T+{\bf R}\cdot {\bf \sigma}=  
\begin{pmatrix}
T + X^3 & X^1 - \imath X^2 \\
X^1+\imath X^2 & T - X^3 
\end{pmatrix}, 
\ee
${\bf \sigma}:=\{\sigma_a\}$ being three {\it Pauli matrices}. 
In the procedure, real time-like $T$ and space-like ${\bf R}=\{X^1,X^2,X^3\}$ 
coordinates are expressed through the primary complex coordinates 
$z^0,\bf z$ as follows: 
\be{newcoord}
T = \vert z^0 \vert^2 + {\bf z}\cdot{\bf z}^*, ~~~~{\bf R} = z^0 {\bf z}^* + 
(z^0)^*{\bf z}+ \imath~{\bf z} \times {\bf z}^*,
\ee
whereas the principal (and {\it non-negative (!)}) Minkowski interval (\ref{idnt}) 
completely reproduces its old form (\ref{Minkinv}):
\be{Minknew}
S^2 = \det X = T^2 - {\bf R}^2 \ge 0.
\ee      

It is noteworthy to distinguish between symmetries of the formerly 
induced geometry and the generalized one defined through the mapping (\ref{newcoord}). 
Under the $SO(3,{\mb C})$ rotations (precisely, under the transformations of 
the covering group $SL(2,{\mb C})$) 
\be{cover}
Z \mapsto A Z A^{-1},~~~A \in SL(2,{\mb C})
\ee
the space-time coordinates $X$ do not, generically, transform 
through themselves. When only the transformation matrix $A$ is {\it unitary}, $AA^+=id$, one has 
a proper law for $X$, namely, $X\mapsto A X A^+$ which in the 
considered case (2:1) corresponds to  usual $SO(3)$ rotations of a 3-vector 
$\bf R$. As to {\it boosts}, they have a special status in the scheme and can be 
accomplished (together with transformations of the whole proper 
{\it Lorentz group}) via {\it left shifts} in the $\mb B$-space, 
\be{ltrans}
Z\mapsto  A Z~~ \Rightarrow X \mapsto A X A^+
\ee
which certainly are {\it not} automorphisms of $\mb B$. One can equivalently  
make use of the {\it right invariant} coordinate frame introduced by 
the conjugate mapping 
\be{conjug}
Z \mapsto Z^+Z.
\ee

To conclude, we have presented a pair of bilinear mappings $Z\mapsto Z\times Z$ 
any of which naturally defines effective coordinates of the {\it causal part} of 
the Minkowski-like space $\bf M$. Indeed, under left (right) shifts of a   
$\mb B$ matrix these coordinates undergo Lorentz transformations. However, 
under an arbitrary $\mb B$ automorphism, they do not, generically, preserve 
their structure and should be defined anew after a ${\mb B}$-symmetry 
transformation. Nonetheless, 
the principal Minkowski interval (\ref{Minknew}) is evidently invariant 
under {\it any} $\mb B$ automorphism (\ref{cover}). 

For further needs, let us mention here (the details can be found in~\cite{Yad}) 
the assumption on {\it random (complex) time} (that is, on random alteration 
of the evolution $\mb C$ valued parameter) and on the resulting random alteration of the 
effective coordinates $X$ of all of the material objects. This conjecture makes 
it possible to identify the increments of effective coordinates $\delta X = 
dZ dZ^+$ with their differences $\Delta X = (Z+dZ)(Z+dZ)^+ - ZZ^+$, by virtue 
of cancellation of the ``interference term''. As a result, at a ``macroscopic'' 
scale one is allowed to regard the {\it bilinear and thus non-holonomic} 
real space-time coordinates $X$ as {\it effectively holonomic} and their 
increments thus as (effectively) {\it full differentials}. It is especially remarkable that 
this very hypothesis, in account of positive definiteness of the effective time 
coordinate (\ref{newcoord}), could resolve the eternal problem of {\it 
time irreversibility}. Indeed, any sequence of {\it random(!)} changes of the 
primary complex coordinates $Z$ necessarily results in an {\it increase} of 
the physical time parameter $T$ ``in average'', $\overline{\Delta T}\ge 0$. 
This means also that one should distinguish between the two time scales, 
the microscopic $T_{mic}$ and averaged macroscopic $T_{mac}$ ones, which are 
related, as it usually takes place in random processes, as $T_{mac}\sim\sqrt 
{T_{mic}}$.

We now pass to a brief review of {\it kinematics} of particle-like formations in 
the primordial complex and induced real spaces. 

\section{Complex null cone, duplicons and the concept of dimerous electron}

Fundamental physical dynamics takes place in the primordial complex 
$\mb B$ space and originates from the solutions of the Cauchy-Riemann-like 
equations generalized to the noncommutative $\mb B$ algebra. Corresponding  
``$\mb B$-differentiable'' functions are considered as distributions of 
fundamental physical field (closely related to twistor or 
2-spinor types of fields) 
while, geometrically, these give rise to congruences of 
(complex or induced real) shear-free rectilinear null rays~\cite{Newman}. 
Singularities (caustics) can be identified with particles and indicate 
their spatial distribution and temporal dynamics.  

Remarkably, as we are going to demonstrate below, complex kinematics is 
nontrivial even for a single (complex) ``world line'' of a point-like 
particle-singularity ${\wh Z}(\tau),~\tau \in {\mb C}$ corresponding to  
the {\it focal line} of the corresponding congruence of 
complex ``null rays''. 
Specifically, let us write down the equation of (local) {\it complex null cone} (CNC) 
of a ``moving'' point singularity:   
\be{CNC}
D:=\det [Z-{\wh Z}(\tau)] \equiv [z^0-{\wh z}^0(\tau)]^2 - 
[{\bf z}-{\bf {\wh z}}(\tau)]^2 =0, 
\ee
$Z$ being an arbitrary fixed point of the $\mb B$ space. CNC equation (\ref{CNC}) 
is, in fact, the compatibility condition for a set of linear equations 
\be{linset}
[Z-{\wh Z}(\tau)] \xi = 0 \Leftrightarrow  \eta: = Z\xi = {\wh Z}(\tau) \xi
\ee
which introduces a principal 2-spinor $\xi\in {\mb C}^2$ and a projective 
twistor ${\bf W}=\{\xi,\eta\},~~\eta \in {\mb C}^2$  fields 
of the congruence. 
By virtue of the {\it incidence relations}, see the r.h.p. of (\ref{linset}),  
values of twistor field $\bf W$ are preserved along any rectilinear element 
of CNC (\ref{CNC}) connecting the point of (instantaneous) particle's 
location $\wh Z(\tau)$ and the point of ``observation'' $Z$ which two 
are thus mutually ``correlated''. 

Importantly, on the corresponding real space-time $X=X^+ \in \bf M$ 
induced through the above constructed mapping $Z\mapsto X=ZZ^+$ fundamental 
equation of CNC (\ref{CNC}) gives rise to equation of the local 
Minkowski {\it light cone} 
\be{lightcone}
DD^* = \det \left([Z-{\wh Z}(\tau)][Z-{\wh Z}(\tau)]^+\right) \equiv \det 
[X-{\wh X}(\tau)] = (T-{\wh T})^2 - ({\bf R}-{\wh {\bf R}})^2 = 0, 
\ee
or, equivalently, to the familiar {\it retardation equation}~\footnote
{However, this equation differs from the usual one in the random complex 
nature of the evolution parameter $\tau$.}. 
It is noteworthy to remark here that such a direct correspondence between 
CNC and real light cone in the induced $\bf M$ space does not take 
place in the formerly introduced~\cite{Mink,Yad} and described in 
Sec.1 geometry within which arbitrary value of the velocity 
of ``propagation of interaction'' ($v\le c$) is allowed. On the contrary, in the above prresented version 
this is always universal and equal to the speed of light. 

Let us now return to consideration of fundamental equation of CNC (\ref{CNC}). 
Contrary to the real case, for a given point $Z\in {\mb C}^4$ it can have a 
great (countable) number of solutions $\tau=\tau_N(Z)$ any of which 
fixes a location ${\wh Z}(\tau_N)$ of the point particle in question at one and 
the same its ``world line''. All these are correlated with the ``observation point'' 
$Z\Rightarrow X=ZZ^+$ in complex as well as in real space along the elements 
of corresponding null cones, that is, are in a ``light-cone interaction''. 
In~\cite{Dupl}, an analogous set of copies of point particle-like formations 
locating at a single complex ``world-line'' and (instantaneously) contributing 
through a ``light cone field'' (i.e., twistor, spinor field etc.) at a 
fixed space-time point $X=ZZ^+$, has been called the ensemble of {\it duplicons}. Concept of duplicons 
explains by itself the observed identity of the primary elements of matter 
reviving thus the old idea of Wheeler-Feynman~\cite{Feynman} about 
{\it all of the 
electrons  as one and the same particle observed in different positions 
at a single (entangled) world line}.  

However, situation arising in the formalism of $\mb B$ algebrodynamics turns 
out to be much more peculiar. In contrast to the permanently existing 
correlation (via the light cone)  between $Z$ and any of the duplicons ${\wh Z}(\tau_N)$, 
{\it true} ``interaction'' can be conducted only via elementary material agents, 
singularities of the $\mb B$ field, or, equivalently, {\it caustics} of the 
complex null rays' congruence. Apart of the focal line itself, these 
are defined by the condition of {\it multiplicity of roots} of CNC 
equation (\ref{CNC}) which reads:
\be{sing}
D^\p: = \frac{dD}{d\tau} = 0.       
\ee

For an arbitrarily taken ``observation point'' $Z$, set of solutions of 
the joint system of equations (\ref{CNC}) and (\ref{sing}) is, generically, 
empty. Instead, one has to deal with a ``world line of an 
{\it elementary} (point-like) 
observer'' $Z_O(\lambda),~~\lambda \in \mb C$ (see for details~\cite{Dupl,Yad}).  
Then equations (\ref{CNC}), (\ref{sing}) (with corresponding 
exchange $Z \leftrightarrow Z_O(\lambda)$), generically, define 
a discrete set of mutually related values of evolution parameters  
$\{\lambda_N,\tau_N\}$ 
indicating the ``instants'' $\lambda_N$ at which a reception of a caustic-signal  
at the observation point occures (with $\tau_N$ being then the analogue of 
the ``retarded time''). This, however, corresponds to the situation 
when {\it some two of duplicons merge} at the point ${\wh Z}(\tau_N)$; this 
corresponds to a multiple root of the CNC equation. It is  
an easy exercise to demonstrate that the arising caustic-signal 
represents itself a null straight line connecting $Z_O(\lambda_N)$ and ${\wh Z}(\tau_N)$ 
(in the underlying complex $\mb B$ space) or, respectively, a rectilinear light ray 
propagating towards an observation point $X_O$ from corresponding point of 
location of two instantaneously merging duplicons $\wh X$ (in the induced 
real space-time $\bf M$). Note that any caustic line (ray) is a locus of 
{\it branching points} of (generally multi-valued) fundamental $\mb B$ field 
(as well as of associated 2-spinor and twistor fields), whereas associated 
gauge and curvature fields undergo infinite amplification (that is, are singular), 
see, e.g.,~\cite{Qanalys} and references therein.     

Thus, in the presented formalism any duplicon is in a sense 
permanently ``invisible'' for a given ``observer'' unless 
at a discrete set of instants when a light-like signal, a ``quantum'', 
arrives from the point of its merging with another duplicon. It is thus 
impossible to regard a single duplicon as a pre-image of an elementary 
particle, as a truly material object. Instead, one is forced to accept the 
concept of ``dimerous electron''~\cite{Yad}. 

Specifically, one concludes that an ``electron'' not only 
``consists'' of two identical point-like parts -- duplicons -- but does not 
even {\it exist} during the whole time interval between some two subsequent 
merging acts. Such a conjecture on the {\it dimerous} nature of electron could 
seem rather strange and insufficiently grounded from the physical viewpoint but  
is supported by a number of recent experimental observations, in particular, 
on {\it fractal charges}. On the other hand, 
rigid mathematical structure of the biquaternionic algebrodynamics 
unavoidably points just to such an ``exotic'' picture of the World. For a more 
detailed discussion of the conjecture we again refer the reader to~\cite{Yad}. 

\section{Invariant geometrical phase and quantum interference}

It has been already mentioned in Sec.1 that the primary complex 
geometry of $\mb B$ space with $\mb C$ valued invariant not only 
induces an effective Minkowski-like real geometry (via its modulus 
part). It also gives rise to an {\it invariant phase} leading 
thus to the geometry of physical space-time with an additional 
$U(1)$ fiber-like structure. This property is completely preserved 
under the generalization of geometry presented in Sec.2. 

Specifically, the principal complex invariant (\ref{det}) of the 
$SO(3,{\mb C})$ automorphisms of $\mb B$ (and of the $SL(2,{\mb C})$ 
left (right) shifts of elements $Z\in \mb B$) can be represented in 
an ordinary exponential form:
\be{exp}
\Sigma = \det Z = Se^{\imath \alpha} 
\ee
where $S=\vert \Sigma \vert$ is the real non-negative Minkowski-like 
interval (\ref{idnt}) expressible through the effective coordinates 
(\ref{newcoord}), 
\be{intrv}
S = \sqrt{T^2-{\bf R}^2} \ge 0, 
\ee           
and $\alpha \in \mb R$ is the above presented phase invariant  
of the $\mb B$-symmetry transformations. Together with $S$, it forms 
the principal {\it evolution parameter} (``complex proper time'') but the 
order of successive events is indefinite and should be assumed additionally,  
through fixing a particular form of the ``evolution curve'' $\alpha=\alpha(S)$
~\cite{Dupl}.

We are now in a position to transparently explain the phenomenon of quantum 
interference without any appeal to the wave-particle dualism~\cite{Yad}. 
Indeed, suppose that two duplicons merge together radiating  
a signal towards an observer ``Obs'' (``preparation'' of an initial 
state ``In'', Fig.1).  After this, the two duplicons diverge in space and, in 
particular, can pass through different slots of a diffraction grating (a crystal). 
However, the final signal (from an ``electron arrived at a screen'') 
one can get when only the two duplicons merge again at a particular 
point of the {\it complex} space (``Out'' state, Fig.1). Since at the initial and final 
points of merging {\it complex coordinates} should be fixed, the phase lags 
along the world lines of duplicons 1 and 2 can differ only by 
$\Delta \alpha = 2\pi N, ~N\in \mathbb{Z}$.

\begin{figure}
\centering
\includegraphics[width=15cm,height=7cm]{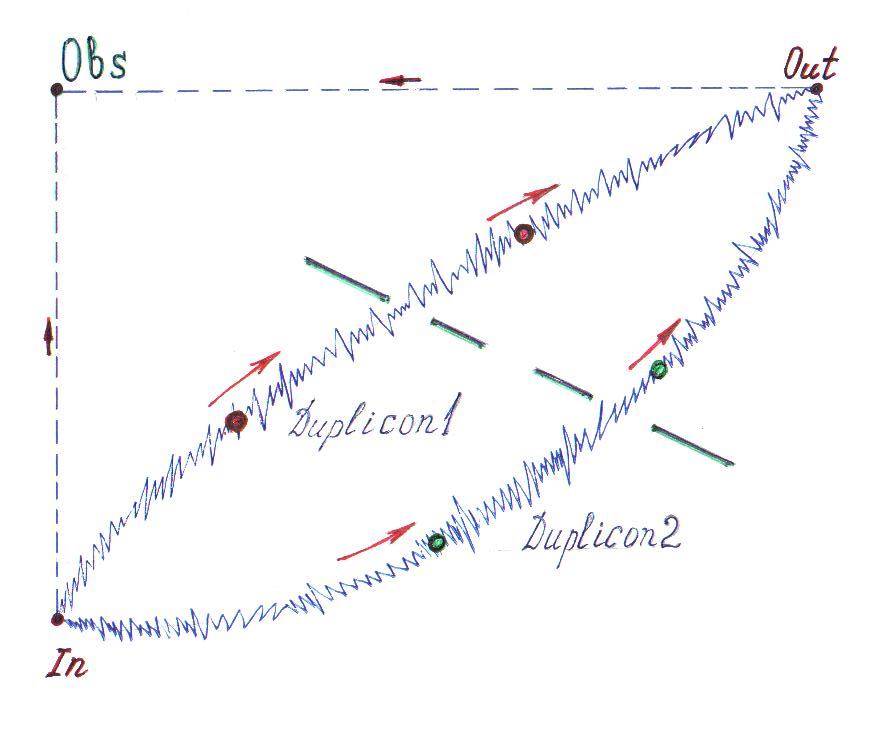}
\caption{\small Mergings of duplicons in a two-slit experiment}
\end{figure}

In the simplest case one assumes that {\it the increment of the geometrical 
phase along any trajectory is proportional to the corresponding increment of 
the proper time}, 
\be{factor}
\delta \alpha = \frac{2Mc^2}{\hbar} \delta S,
\ee
$M$ being the electron rest mass (which aquires here the sense of a universal 
physical constant). We direct the readers' attention that  this relationship 
is, in fact, of universal and fundamental nature. In the forthcoming papers we 
are going to demonstrate this, in particular, on the base of the ideas 
of I.A. Urusovskii and the 6D geometry of extended space-time he proposed 
(see, e.g.,~\cite{Urusovskii} and references therein).

Note also that the proportionality 
factor is taken to be equal to the doubled {\it de Broglie frequency} of an  
hypothetical ``internal gyration'' of a microparticle. Here, however, 
none oscillation actually takes place~\footnote{Just such a doubled value of 
internal frequency naturally arises in a number of alternative approaches, 
in particular, in different models of classical spinning particles~\cite{Rivas,Jena}}, 
and the phase lag is completely of algebro-geometrical nature. 
We notice also that, in the scheme in question, precisely this numerical 
value of effective frequency has been choosed in order to establish 
correspondence with the non-relativistic limit, see below.

Consequently, the (discrete) points 
of possible detection of electrons at the screen exactly correspond  
in position to the condition of maxima of the classical interference, namely 
\be{interfer}
\Delta \alpha = (2Mc^2/\hbar)\Delta \int \delta S = 2\pi N, 
\ee
where $\Delta$ in the r.h.p. means difference of the lengths of the 
duplicons' world lines connecting the points of some two subsequent mergings. 
This is mathematically equivalent to the condition for total change of the 
(non-integrable even after the averaging procedure) 
proper time $\Delta S$ along a corresponding closed loop $1-2-1$, so that one 
obtains the following fundamental {\it condition of relativistic quantum 
interference}:
\be{qinterf}
\frac{Mc^2}{h} \oint \delta S = \frac{N}{2}, 
\ee
which in a sense explains the mystery of closed time loops arising  
in the framework of different attempts of classical interpretations 
of quantum interference phenomena (see, e.g.,~\cite{Rave}). 

In the non-relativistic approximation with respect to the (averaged) velocity 
of duplicons' motion $V: = \vert \delta {\bf R}/\delta T\vert << 1$, one has 
\be{approx}
\delta S = \sqrt{\delta T^2 - \delta {\bf R}^2} = \delta T 
\sqrt{1-\frac{V^2}{c^2}}
\approx \delta T (1-\frac{V^2}{2c^2}) = \delta T -\frac {V}{2c^2} \delta L, 
\ee
where $\delta L := V \delta T$ is the increment of the (averaged) path length of 
a duplicon in the 3D real physical Euclidean space. Taking in 
account that 
the increment of an averaged time interval $\delta T$ itself may be effectively 
considered as a full diferential (see the end of Sec.2), in the first order 
approximation in $V/c << 1$ condition of ``quantum interference'' aquires the form
~\cite{Yad}
\be{nonrel}
\oint \frac {\delta L}{\Lambda} = N,~~~\Lambda: = \frac{h}{MV},   
\ee
so that one concludes in the non-relativistic approximation:
                                           
\vskip2mm
\noindent
{\it A pair of duplicons may undergo two subsequent mergings (when only they 
radiate light-like signals and can thus be detected and identified as a whole 
``electron'') at the points for which the (averaged, or smoothed) length of 
the closed loop formed by their 3D trajectories is integer in fractions of the 
de Broglie wavelength of electron $\Lambda$}.   

\vskip2mm       
  
\begin{figure}
\centering
\includegraphics[width=10cm,height=5cm]{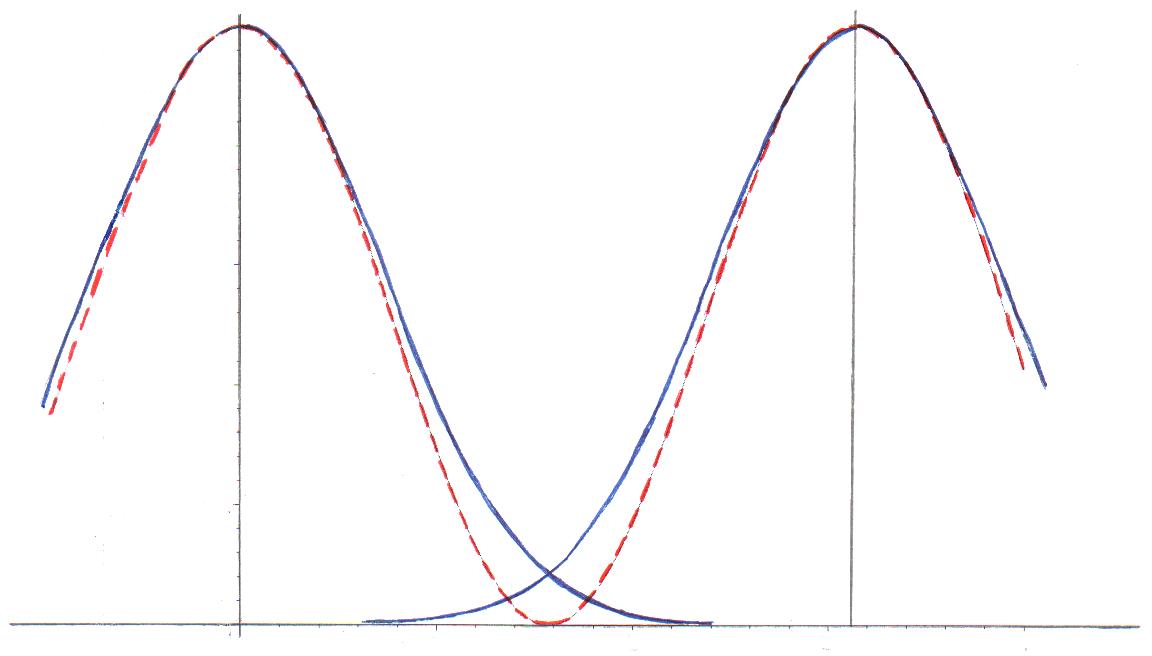}
\caption{\small Distribution of probabilities in the two-slit 
experiment, according to algebrodynamics (solid line) and quantum theory 
(dotted lines).}
\end{figure}

Note that in the {\it free case} the above exposed procedure 
seems to be very close to the Feynman's representation for the 
wavefunction of a self-interfering micro-particle. 
However, here we do not appeal to the concept of the 
probability amplitude and even do not assume any wave-like 
properties of the matter dealing instead with the notion of the phase of 
a purely geometrical nature. A simple interference expirement to 
distinguish between the predictions of quantum probabilistic theory and 
the algebrodynamical scheme can be suggested. 

Specifically, only in an {\it idealized situation} there exists a discrete 
set of points (at a screen) where the electrons (represented by merging 
duplicons) could be detected. In account of {\it statistical errors}, 
however, one would observe a {\it Gauss-like} distribution of probabilities 
\be{gauss}
w\sim \exp{\{-\frac{(x-x_N)^2}{l^2}\}}
\ee 
in the vicinity of any of such points 
(with coordinates of the center $x_N$ and dispersion $l$), see Fig.2, solid   
lines. At the same time, quantum theory predicts the interference pattern with 
maxima coinciding with $x_0$ and distribution of probabilities determined by 
the wave-like propagation of amplitudes and represented by the function
\be{cos}
w\sim \cos^2\{\frac{x-x_N}{\Lambda}\} 
\ee
(in the non-relativistic approximation), see Fig.2, 
dotted line. If the dispersion value is about that of the de Broglie 
wavelength, $l\sim \Lambda$ (what is just a necessary condition for 
diffraction phenomenon to be observed), the predicted distributions coincide 
near maxima (up to the second derivatives) and are very close globally (Fig.2). 
Special consideration is thus necessary to distinguish the predicted 
distributions in the course of a standard electron diffraction  experiment; 
its details will be discussed elsewhere.  

\section{Conclusion}

In the article we reproduce the main results of $\mb B$ algebrodynamics 
on the base of the proposed general correspondence between the 
primordial complex geometry 
and (phase extension of) real physical space-time geometry (Sec.2). 
It was shown that the formerly introduced concept of an ensemble of identical 
point-like objects, duplicons, does not seem to explicitly represent the  
real matter pre-elements, say, electrons (in the spirit of the   
famous ``one-electron Universe'' of 
Wheeler-Feynman). The true correspondence turns out to be much more refined and 
manifests itself at the instants of merging of some two of the duplicons, when 
a light-like signal is radiated towards an observation point and 
the ``electron'' can only be detected.  

Such interpretation allows for transparent and successive explanation of the 
standard two-slit experiment, without invoking any quantum mechanical  
formalism and the probability amplitude paradigm in particular, though in 
some aspects it resembles the Feynman path-integral treatment. Moreover, 
the obtained relativistic condition for the location of 
``interference maxima'' (\ref{qinterf}) is a direct generalization of the 
familiar de Broglie non-relativistic relation (\ref{nonrel}) and {\it 
must be taken in account even in the framework of the orthodox quantum theory}. 
Indeed, formula (\ref{qinterf}) seems to be an explicit relativistic generalization 
of the {\it Bohr-Zommerfeld quantization condition} for periodic motion and here, 
moreover, it follows just from first principles.  
As for successive algebrodynamical approach, it only 
slightly differs in predictions of the probability distribution from those 
of the quantum theory; nonetheless, the difference could be experimentally revealed. 

We conceive, of course, that the classical-geometrical explanation of a single 
quantum phenomenon is insufficient for seriously taking the approach as a 
consistent alternative to the accepted quantum paradigm. However, 
the above presented treatment visually demonstrates that the misterious 
quantum phenomena might receive 
quite unexpected and even striking explanation on the base of pure 
geometry. We hope, moreover, that other phenomena including ``quantum 
non-locality'' will also find a clear classical interpretation in this 
framework.

\end{document}